\documentclass[thmsa,12pt]{article}
\usepackage{amssymb}

\usepackage{sw20lart}

\input{tcilatex}
\input tcilatex

\begin{document}

\date{}
\title{BOSE\ - EINSTEIN\ PARTITION\ STATISTICS\ OF\ PHOTONS\ EMITTED\ FROM\ A\
SUPERRADIANT\ ACTIVE\ MICROCAVITY}
\author{E. De Angelis, F. De Martini, and P. Mataloni \\
Dipartimento di Fisica and \\
Istituto Nazionale di Fisica della Materia\\
Universit\`{a} di Roma ''La Sapienza'', Roma, 00185 - Italy}
\maketitle

\begin{abstract}
We report the results of the first investigation on the superradiant
temporal and spatial quantum dynamics of two dipoles excited in a planar
symmetrical microcavity by a controlled femtosecond two-pulse excitation. A
superradiant enhancement of the time decay of the dipole excitation for a
decreasing inter-dipole transverse distance $R$ has been found. Furthermore,
the photon partition statistics of the emitted field is found to exhibit a
striking quantum behaviour for $R\leq l_{c}$, the transverse extension of
the single allowed microcavity mode.

\end{abstract}

It has been demonstrated that the planar active microcavity behaves as a
reliable source of nonclassical radiation when small ensembles of
fluorescent molecules are isolated within the optical resonator and excited
by a sequence of femtosecond laser pulses \cite{1,2}. In these conditions
the ultrashort pulse excitation forbids the dynamical recycling between the
lower and the excited molecular levels leading to a strictly single photon
emission over one of the two allowed external output modes, \textbf{k} and%
\textbf{\ k', }of the microcavity with relevant ''microscopic'' dimension $%
d=m\frac{\lambda }{2}$ and $m=1$. We can double this process by focusing two
independent femtosecond pulses, within a transverse focal spot dimension $%
\sim \lambda _{p}$, the excitation wavelength (wl), to excite two dipoles
located within the microcavity at a mutual transverse distance $\mathbf{R}$,
on the symmetry plane \textbf{Z }= 0. The problem of the transverse quantum
interaction between two dipoles in a cavity, generally with different
spatial orientations, has been so far only theoretically addressed in terms
of superradiance in atomic spontaneous emission (SE) \cite{3,4}. In recent
work we have found that a typical superradiant regime is established in the
microcavity when $R\equiv \left| \mathbf{R}\right| \mathbf{\ }$is shorter
than $\ell _{c}=2\lambda \sqrt{fm}$, the ''transverse coherence length'',
i.e., the effective radius of the gaussian - like e.m. mode in the cavity,
being $\lambda \ $the emitted wavelength and\ $f\gg 1$ the cavity
''finesse'' \cite{5}. In particular we found experimentally that the
dynamics of the two dipoles within the microcavity is causally connected
within a retardation time $\tau \lesssim \ell _{c}/c$ which is in the
sub-picosecond time i.e., corresponding precisely to the time resolution
allowed by a standard femtosecond laser technique \cite{6,7}. It is worth
noting in this connection that in \textit{free-space}, i.e., in absence of
cavity confinement, the corresponding two-atom superradiant coupling can be
established only over a ''microscopic'' inter-atomic transverse distance $%
R\approx $ $\lambda $. This makes impossible the selective, independent
optical excitation of the interacting fixed dipoles as well as any
controlled photon emission over a limited number of modes \cite{3,5}.
Furthermore, the peculiar, highly favorable topological properties of the 
\textit{planar} microcavity in superradiance investigations are further
enlightened by considering that the adoption of the spherical or confocal
resonators, of general use in modern cavity QED, again is associated with a
''microscopic'' size of $\ell _{c}$ thus reproducing closely the free-space
impossibility condition. In our present experiment two laser beams were
focused by a common lens with f.l. $20$ $cm$ on the active plane of a single
mode symmetrical microcavity in two focal spots with diameter $\varphi =10$ $%
\mu m\ $at an externally adjustable mutual transverse distance $\mathbf{R\ }$%
along the spatial \textbf{Y} axis, (cfr: Figure 1). The corresponding two
excitation optical pulses, generated by an amplified colliding pulse
mode-locking (CPM) dye laser with wl $\lambda _{p}=615$ $nm\ $and duration $%
\delta t=80$ $fs$. The active medium consisted of a $10^{-5}M/\ell $
concentration of Oxazine 725 molecules in a matrix of polymethyl
methacrylate (PMMA) solid film, cooled at liquid nitrogen temperature. The
wl of the emitted radiation was $\lambda =700$ $nm.$ The microcavity
consisted of a single-longitudinal mode Fabry-Perot interferometer,
terminated by two parallel, plane multilayer dielectric equal mirrors,
highly reflecting ($\mathcal{R}\equiv \left| r\right| ^{2}=.9990$) at $%
\lambda $ and highly transparent ($T=.98)\ $at $\lambda _{p}$. The cavity
''finesse'' was $f=3000$. This value determines the microcavity storage time
which is equal to the ''coherence time'' of the emitted particles: $\tau
_{c}\approx 1$ $ps$. The condition of single-photon emission from the
microcavity following any single laser pulse excitation, indeed a critical
condition in the context of the present work, was carefully tested
experimentally by use of suitable Hanbury Brown-Twiss (HBT) interferometric
configurations involving each or, alternatively, both output modes, \textbf{k%
} and\textbf{\ k' }\cite{1}\textbf{. }The HBT coincidence rate, evaluated as
the ratio between the number of spurious two-detector coincidences and the
number of detected ''singles'' per second, was found less than $10^{-3}$.
The output photons were detected by cooled, avalanche single photon-counting
modules EGG-SPCM200. They are indicated by $D_{1}$, $D_{2}$, $D_{3}$ in
Figure 1. The typical quantum efficiency of the three equal detectors was $%
70\%$. The significant time measurements were carried out by feeding a
time-to-amplitude converter (TAC) with the standard TTL output pulses of
couples of detectors. The TAC (Silena 7412) was connected to a Multichannel
Analyzer (MCA)\ (Silena 7923-2048).\ The output radiation, spectrally
filtered by the microcavity within a bandwidth $\Delta \lambda =0.2$ $nm$,
was focused into the active surface of the detectors, with diameter $=100$ $%
\mu m$, by $5$ $cm$ f.l.\ lenses. Because of the random orientation of the
active molecules the output radiation was found slightly ($20\%$) linearly
polarized along the\ (linear) polarization of the excitation laser beams 
\cite{8}. In order to investigate the emission properties of the active
dipoles along the orthogonal transverse spatial directions \textbf{X} and 
\textbf{Y}, the output radiation detected by each $D_{j}\ $was filtered by
adjustable optical polarization analyzers $P_{j}$. The polarization of the
excitation pulses was set oriented along \textbf{X}. Two different
experimental configurations, involving two laser pulse excitation, were
investigated. Configuration A) in Figure 1: by adoption of $D_{1}$ and $%
D_{2} $ as \textit{start} and \textit{stop} devices for the TAC, we could
measure photon pairs emitted over the single output mode \textbf{k, }i.e%
\textbf{. }on one side of the microcavity. Configuration B): the adoption of 
$D_{1}$ and $D_{3}$ as \textit{start-stop} devices, allowed direct HBT
investigations on both output modes \textbf{k} and \textbf{k' }of the
microcavity, here used as a kind of \ \textit{active beam splitter} \cite{1}.

The first experiment, involving the configuration A) in Figure 1, consisted
of the measurement of the temporal evolution of the second-order electric
field correlation function $F(\tau )=<\widehat{\mathbf{E}}^{-}(t)\widehat{%
\mathbf{E}}^{-}(t+\tau )\widehat{\mathbf{E}}^{+}(t+\tau )\widehat{\mathbf{E}}%
^{+}(t)>\propto P_{\mathbf{k},\mathbf{k}^{\prime }}(2,0)$, where $\tau $
represents the time difference between the output pulses released by the two
detectors and $P_{\mathbf{k},\mathbf{k}^{\prime }}(2,0)\ $is the probability
of detection of 2 photons over the mode \textbf{k} and zero photons over the
mode \textbf{k}'. The inset of Figure 2 shows two typical, normalized MCA
curves obtained for two extreme values of the inter-dipole spacing, $R$. The
solid curve corresponds to $R=0.33\ell _{c}=25$ $\mu m$, being $\ell _{c}=77$
$\mu m$\ for our experiment. The dotted curve corresponds to $R=7.2\ell
_{c}=570$ $\mu m$. In Figure 2 we have reproduced the same results in a
semilog scale, for small values of $\tau $, by adding a further intermediate
set of data, taken for $R=2.9\ell _{c}=230$ $\mu m$. All these results have
been obtained with both polarizers $P_{j}$ oriented along a common spatial
direction \textbf{X. }The experimental results show a marked enhancement, of
a factor $1.8$, of the rate of spontaneous emission (SE)\ by two interacting
dipoles which are selected to be mutually parallel and placed at a mutual
distance $R\lesssim \ell _{c}$. The enhancement effect has been found to
disappear when $P_{1\ }$and $P_{2}$ are set mutually orthogonal, i.e., along
the directions \textbf{X} and \textbf{Y}. In this case the rate of
spontaneaous emission is found equal to the SE value $\Gamma _{\infty
}\equiv T^{-1}$, of a single atom in the microcavity \cite{9} and to the
lowest value obtained in the previous case, i.e., for $R>>\ell _{c}$. In
addition, when both analyzers $P_{j}$ are oriented along the \textbf{Y }%
axis, corresponding to the less efficient head-on dipole-dipole interaction,
the SE enhancement effect has been found to be reduced of about a factor $%
1.2 $. In summary, all these results realize the expected features of a
superradiant model involving two dipoles e.m. interacting in an optical
microcavity.

In order to get a better insight into this process, let us consider here the
simple case of the time evolution of the field radiated by two equal
dipoles, both parallel to the \textbf{X} axis, excited instantaneously at
the time $t_{0}=0$ and observed at a later time $t$ by a detector located on
the \textbf{Z }axis at a distance $Z>>\lambda $ from the center of a
symmetrical, lossless microcavity. In the Heisenberg representation, the
field$\ $can simply be expressed in terms of the transition operators $%
\widehat{\mathbf{\pi }}_{A}(t)\ $and $\widehat{\mathbf{\pi }}_{B}(t)\ $of
the two dipoles $A$ and $B$ \cite{9,10}:\smallskip

\begin{equation}
\widehat{\mathbf{E}}^{+}(Z,t)=-\Theta \left( 1+r\right) (1-\left| r\right|
^{2})^{\frac{1}{2}}\left[ \widehat{\mathbf{\pi }}_{A}(t-\frac{Z}{c})+%
\widehat{\mathbf{\pi }}_{B}(t-\frac{Z}{c})\right] \stackrel{\infty }{%
\stackunder{n=0}{\sum }}r^{2n}
\end{equation}
where the effect of multiple intracavity reflections is considered, $r$ is
the value of reflection coefficient of the mirrors at normal incidence and $%
\Theta $ is a constant. We may insert the above expression into the
definition for $F(\tau )$, by assuming that the atomic operators labelled by 
$A$ and $B$ mutually commute at first order and by taking into account only
the terms in which each $\widehat{\mathbf{\pi }}\ $operator is multiplied by
a $\widehat{\mathbf{\pi }}^{\dagger }$ corresponding to either atom $A$ or $%
B $ \cite{10}. We may further make use of the ansatz $\widehat{\mathbf{\pi }}%
(t)=\widehat{\mathbf{\pi }}(0)\exp \left[ -\left( i\frac{2\pi c}{\lambda }+%
\frac{1}{2}\Gamma (R)\right) t\right] \ $\ implying that no causal
inter-dipole interaction is established at $t_{0}=0$ \cite{3,7}. At last,
the second order correlation function may be written as: $F(\tau )\propto $ $%
\ \stackunder{i}{\sum }\left\langle \widehat{\mathbf{\pi }}_{i}^{\dagger }(t)%
\widehat{\mathbf{\pi }}_{i}^{\dagger }(t+\tau )\widehat{\mathbf{\pi }}%
_{i}(t+\tau )\widehat{\mathbf{\pi }}_{i}(t)\right\rangle $\ \ + $\ 
\stackunder{i\neq j}{\sum }\left[ \left\langle \widehat{\mathbf{\pi }}%
_{i}^{\dagger }(t)\widehat{\mathbf{\pi }}_{j}^{\dagger }(t+\tau )\widehat{%
\mathbf{\pi }}_{j}(t+\tau )\widehat{\mathbf{\pi }}_{i}(t)\right\rangle
\right. $ \ + $\left. \left\langle \widehat{\mathbf{\pi }}_{i}^{\dagger }(t)%
\widehat{\mathbf{\pi }}_{j}^{\dagger }(t+\tau )\widehat{\mathbf{\pi }}%
_{i}(t+\tau )\widehat{\mathbf{\pi }}_{j}(t)\right\rangle \right] $, for $i$, 
$j=A$,$B$. The first sum vanishes owing to the antibunched character of the
emitted radiation. By replacing in the second sum the ensemble averages with
time averages, the expected result is found by a simple integration: $F(\tau
)\varpropto \exp \left( -\Gamma (R)\left| \tau \right| \right) $. This
result may be compared with the experimental MCA\ output data reported in
the inset of Figure 2. The explicit expression of $\Gamma (R)$, evaluated by
a fully relativistic quantum field theoretical analysis, is expressed as a
function of the \textit{free-space} SE rate $\gamma =\frac{1}{2}(T\QATOP{-1}{%
SE})\ $of a $\sin $gle dipole \cite{3}. In the case of coupled dipoles
commonly oriented along \textbf{X}\ the following result is found:

\begin{eqnarray}
\Gamma (R) &=&\gamma \left\{ 1+\frac{3}{k^{3}}{\LARGE [}\sin \left(
kR\right) {\LARGE (-}\frac{1}{R^{3}}+\frac{k^{2}}{R}{\LARGE )}+\cos \left(
kR\right) \frac{k}{R^{2}}{\LARGE ]}\theta \left( ct-R\right) +\right. 
\nonumber \\
&&+\frac{3}{k^{3}}\stackrel{\infty }{\stackunder{n=1}{\sum }}\left( -\left|
r\right| \right) ^{n}\cdot \left\{ {\LARGE [}\sin \left( knd\right) {\LARGE (%
}-\frac{1}{\left( nd\right) ^{3}}+\frac{k^{2}}{nd}{\LARGE )}+\right. 
\nonumber \\
&&+\cos \left( knd\right) \frac{k}{\left( nd\right) ^{2}}{\LARGE ]}\theta
\left( ct-nd\right) +{\LARGE [}\sin \left( kR_{n}\right) {\LARGE (}-\frac{1}{%
R_{n}^{3}}+\frac{k^{2}}{R_{n}}{\LARGE )}+  \nonumber \\
&&\left. \left. +\cos \left( kR_{n}\right) \frac{k}{R_{n}^{2}}{\LARGE ]}%
\theta \left( ct-R_{n}\right) \right\} \right\}
\end{eqnarray}
where $R_{n}=\sqrt{R^{2}+\left( nd\right) ^{2}\text{ }}$ and $\theta \left(
ct-x\right) \ $are Heaviside step functions accounting for relativistic
causality in the establishment of the inter dipole interactions. By adopting
the actual values of the microcavity parameters within an explicit numerical
evaluation, it is found that the value of $\Gamma (R)$ corresponding to the
condition of maximum superradiance ($R=0$) is \textit{twice }as large as the
value of the case of two independent dipoles ($R>>\ell _{c}$).\ This is in
good agreement with the experimental results for SE time decay, as shown by
Figure 2.

The above results show once again that the peculiar\textit{\ }topology of
the microcavity is instrumental in the determination of the \textit{time}
behavior of a quantum SE decay process within an inter-atomic interaction.
It is easy to recognize that the \textit{mesoscopic} character of the device
is precisely ascribable to the fact that the De Broglie wavelength $\lambda
\ $of the confined particle, the photon, is of the order of the relevant
dimension $d$ is of the confining device. This is a common characteristics
of all nanostructures that exhibit quantum properties. In this perspective,
it is expected that also the \textit{spatial} behaviour of some relevant
dynamical process should be affected by the peculiar quantum properties of
the device.

We may look for instance at the spatial statistical distribution of the
couples of photons emitted over the two allowed microcavity output modes%
\textbf{\ k} and \textbf{k'} under corresponding couples of excitation laser
pulses. This process has been investigated with the same microcavity by both
experimental configurations A) and B), Figure 1, and for very small time
delay $\tau \approx 0$. Precisely, we have measured the probability $P_{%
\mathbf{k},\mathbf{k}^{\prime }}(2,0)\ $of the simultaneous phodetections
realized by $D_{1}$ and $D_{2}\ $coupled to one external output mode \textbf{%
k}, and the probability $P_{\mathbf{k},\mathbf{k}^{\prime }}(1,1)\ $of the
simultaneous photodetections realized by\textbf{\ }$D_{1}$ and$\ D_{3}$
coupled to the counterpropagating external output modes (\textbf{k,} \textbf{%
k'). }By assuming a ''classical'' Maxwell-Boltzmann partition statistics,
and by accounting for the couples of detection events, we expect: $P_{%
\mathbf{k},\mathbf{k}^{\prime }}(2,0)=1/3$, $P_{\mathbf{k},\mathbf{k}%
^{\prime }}(1,1)=2/3$. This implies that the value of $\Gamma (R)$ measured
by the experimental configuration B) is twice as large as the one measured
by configuration A). The experimental results given in Figure 3 show that
this is indeed verified for a large inter-dipole distance: $R>>$ $\ell _{c}$%
. However, these results also show that, for shorter distances $R\lesssim $ $%
\ell _{c}$, the relative values of the probabilities converge toward the
common value: $P_{\mathbf{k},\mathbf{k}^{\prime }}(1,1)=P_{\mathbf{k},%
\mathbf{k}^{\prime }}(2,0)=1/2$. This implies that a quantum Bose-Einstein
(BE)\ partition process determines the photoemission over the external modes
of the microcavity, a device that is then found to behave like a two photon
''quantum lamp''. The realization of a quantum statistical photon
distribution law at the output of an optical cavity has never been
investigated before. We may try to explain this remarkable quantum
phenomenon by the following simplified argument. Because of the condition $%
R\ll $ $\ell _{c}$, the two photons are emitted over the allowed stationary
mode of the microcavity, which consists of the superposition of the two
travelling - wave modes associated with the \textit{internal }momenta $\hbar 
\mathbf{k}$ and $\hbar \mathbf{k}^{\prime }=-\ \hbar \mathbf{k}$. Since the
cavity mode identifies the spatial extent of the minimum - uncertainty
application of the Heisenberg principle of the tridimensional photon
dynamics, the two particles are \textit{in principle }dynamically
''indistinguishable'' and then the \textit{internal} two photon state is
expressed in terms of the momentum eigenvectors by: $\left| \Psi
\right\rangle _{in}=3^{-\frac{1}{2}}(\left| 2,0\right\rangle +\left|
1,1\right\rangle +\left| 0,2\right\rangle )$, where: $\left|
x,y\right\rangle \equiv $ $\left| x\right\rangle _{k}\left| y\right\rangle
_{k^{\prime }}$. Note that, according the the BE\ partition law, the state $%
\left| 1,1\right\rangle \ $is counted only \textit{once} within the
structure of $\left| \Psi \right\rangle _{in}$. Consider now that the two
photon transmission throught the mirrors of the lossless microcavity
formally corresponds to a simple \textit{unitary} transformation leading to
: $\left| \Psi \right\rangle _{out}\propto $ $\left| \Psi \right\rangle _{in}
$. In the expression of $\left| \Psi \right\rangle _{out}$, the function
accounting for the two photon state \textit{outside} the cavity, $\hbar 
\mathbf{k\ }$and $\hbar \mathbf{k}^{\prime }=-\ \hbar \mathbf{k\ }$are to be
interpreted as \textit{external} momenta, i.e. affecting the particle
dynamics outside the cavity. This leads to the quantum result found
experimentally for $R\ll $ $\ell _{c}$: $P_{\mathbf{k},\mathbf{k}^{\prime
}}(1,1)\equiv \left| \left\langle 1,1\right| \left. \Psi _{out}\right\rangle
\right| ^{2}=P_{\mathbf{k},\mathbf{k}^{\prime }}(2,0)\equiv \left|
\left\langle 2,0\right| \left. \Psi _{out}\right\rangle \right| ^{2}=1/2$.
On the other hand, the couple of photons created by two distant dipoles, $%
R\gg $ $\ell _{c}$ do not belong to the same\ gaussian - like internal
stationary microcavity mode and then the above indistinguishability
condition is lost. As this condition is reproduced outside the cavity within
the detection process it leads eventually to the realization of the
classical result $P_{\mathbf{k},\mathbf{k}^{\prime }}(1,1)=2P_{\mathbf{k},%
\mathbf{k}^{\prime }}(2,0)$. In summary we have found that, for $R/\ell
_{c}\ll 1$, the two photons tend to be emitted at the \textit{same} time and
over the \textit{same} spatial output mode of the microcavity. This striking
spatio - temporal ''photon dragging'' process, found here under a strictly
controlled simultaneous two-photon emission, could also be detected by
exciting a large unknown number of active molecules in the single mode
microcavity. In this case the quantum character of the multiphoton
statistical process can be identified by the experimental determination of
the two-channel ''quantum noise function'' introduced in a different context
by work \cite{11}. A more detailed description of the experiment and a more
extended theoretical analysis will be given in a following paper. This
research was carried out under the CEE-TMR Contract ERBFMRXCT96-0066. We
also thank MURST and INFM (Contract No PRA97-cat) for funding.

\textbf{FIGURE CAPTIONS}

\begin{description}
\item[Fig. 1 - ]  Optical configurations A) and B) of the
Hanbury-Brown-Twiss interferometers. In both configurations the couples of
equal single photon detectors are connected to a Multichannel Analyzer (MCA)
through a Time-to-Amplitude Converter (TAC).

\item[Fig. 2 - ]  Two photon correlation function $F\left( \tau \right) $,
in semilog scale, as function of the delay $\tau \ $between the emitted
photons for different relative values of the transverse inter-dipole
distance: $R/\ell _{c}=0.33$ ($\mathbf{\bigtriangleup }$), $2.9$ ($\circ $),
and $7.2$ ($\mathbf{*}$). The dashed straight lines represent the
corresponding time decays evaluated by a quantum analysis. Inset:\
experimental normalized MCA distributions $F\left( \tau \right) $ for $%
R/\ell _{c}=0.33$ (continuous line) and $R/\ell _{c}=7.2$ (dotted line).

\item[Fig. 3 - ]  Two photon partition probabilities $P_{\mathbf{k},\mathbf{k%
}^{\prime }}(1,1)$ and $P_{\mathbf{k},\mathbf{k}^{\prime }}(2,0)$ over the
two channels \textbf{k}, \textbf{k'}, detected at $\tau \approx 0\ $as
function of the relative transverse inter-dipole distance $R/\ell _{c}$. \
Note that the common quantum Bose - Einstein value of the probabilities and
the two different classical Maxwell-Boltmann (M-B) values are reached
correspondingly for $R/\ell _{c}\ll 1\ $and $R/\ell _{c}\gg 1$.
\end{description}

\end{document}